\documentclass[aps,preprint,superscriptaddress]{revtex4}

\usepackage{natbib}
\usepackage[dvips]{graphicx}
\usepackage{amsmath}
\usepackage{amssymb}
\usepackage{ifpdf}

\newcommand{\pz}{p_{e\parallel}}

\newcommand{\pp}{p_{e\perp}}
\newcommand{\grl}{\emph{Geophys. Res. Lett.}}

\newcommand{\bhat}{{\bf\hat{b}}}

\begin{document}

\title{Three-Dimensional Stability of Current Sheets Supported by Electron Pressure Anisotropy}

\author{A. Le}
\affiliation{Los Alamos National Laboratory, Los Alamos, New Mexico 87545, USA}
%\affiliation{Space Science Institute, Boulder, CO 80301, USA}

\author{A. Stanier}
\affiliation{Los Alamos National Laboratory, Los Alamos, New Mexico 87545, USA}

\author{W. Daughton}
\affiliation{Los Alamos National Laboratory, Los Alamos, New Mexico 87545, USA}

\author{J. Ng}
\affiliation{University of Maryland, College Park, Maryland 20742, USA}

\author{J. Egedal}
\affiliation{University of Wisconsin---Madison, Madison, Wisconsin  53706, USA}

\author{W. D. Nystrom}
\affiliation{Los Alamos National Laboratory, Los Alamos, New Mexico 87545, USA}

\author{R. Bird}
\affiliation{Los Alamos National Laboratory, Los Alamos, New Mexico 87545, USA}

\begin{abstract}

The stability of electron current sheets embedded within the reconnection exhaust is studied with a 3D fully kinetic particle-in-cell simulation. The electron current layers studied here form self-consistently in a reconnection regime with a moderate guide field, are supported by electron pressure anisotropy with the pressure component parallel to the magnetic field direction larger than the perpendicular components, and extend well beyond electron kinetic scales. In 3D, in addition to drift instabilities common to nearly all reconnection exhausts, the regime considered also exhibits an electromagnetic instability driven by the electron pressure anisotropy. While the fluctuations modulate the current density on small scales, they do not break apart the general structure of the extended electron current layers. The elongated current sheets should therefore persist long enough to be observed both in space observations and in laboratory experiments.    

\end{abstract}

\maketitle

\section{Introduction}
By breaking the so-called frozen-in condition of ideal MHD, magnetic reconnection opens a pathway for the release of stored field energy in space, astrophysical, and laboratory plasmas \cite{priest:2002}. In collisionless regimes, the details of the small-scale layers where reconnection occurs, typically referred to as diffusion regions, depend on electron and ion kinetic effects. Understanding reconnection regions down to ion and electron kinetic scales is becoming even more important as substantial observational and experimental efforts have been undertaken to gather real-world data resolving the kinetic scales. Measuring the electron kinetic scales was a main goal of NASA's Magnetospheric Multi-Scale (MMS) mission \cite{moore:2013}, and electron diffusion region data has been gathered from both the magnetopause \cite{burch:2016} and the magnetotail \cite{torbert:2018}. In addition, the Terrestrial Reconnection Experiment (TREX) at the University of Wisconsin was designed specifically to access the nearly collisionless regime \cite{le:2015}, with data revealing new electron-scale kinetic processes \cite{olson:2016}.

The interpretation of measurements of reconnection regions relies heavily on comparing data to the signatures derived from kinetic numerical models (e.g., \cite{hesse:1999,pritchett:2001,fujimoto:2006,daughton:2006,ng:2011,shuster:2015,shay:2016,bessho:2016,egedal:2016,egedal:2018}). In recent work, we found that the details of the reconnection region, including both the electron diffusion region and the reconnection exhaust out to ion scales, depend sensitively on the parameters of the numerical simulations \cite{le:2013}. The qualitative features of the reconnecting current sheets fall into different regimes depending on whether the thermal electrons follow magnetized, meandering, or chaotic orbits \cite{buchner:1989}. The classes of electron orbits in turn depend on the electron pressure normalized to the magnetic pressure $\beta_e$, the value of the out-of-plane guide magnetic field, and the ion-to-electron mass ratio $m_i/m_e$ employed in the simulation. When high mass ratios of at least several hundred are used, a new regime ("Regime 3" of Ref.~\cite{le:2013}) opens for a range of moderate guide magnetic fields $B_g/B_0\sim$0.15---0.6 (with $B_0$ the upstream reconnecting field component). This regime includes an elongated electron current sheet supported by pressure anisotropy that extends many ion inertial lengths into the reconnection exhaust. The regime spans a range of parameters typical of magnetospheric plasmas, and we expect the extended electron current sheets to develop naturally in the magnetotail.

Because the regime of extended anisotropy-supported electron current sheets requires a relatively large value of $m_i/m_e$, it was previously studied only in 2D kinetic simulations. This begs the question of whether the elongated current sheets persist in real 3D systems, or whether, for example, they break apart into flux ropes like current sheets that develop along the magnetic separatrices \cite{daughton:2011}. The question of 3D stability is particularly important for determining whether the embedded elongated electron current sheets could be observed in spacecraft data or laboratory experiments. Here, we address this question with a 3D fully kinetic simulation. The simulation suggests that the elongated electron current sheets are modulated by several short wavelength instabilities, including lower hybrid drift modes and an electromagnetic mode driven by the electron pressure anisotropy. The overall current sheet structures are nevertheless fairly robust and should remain coherent long enough to be observed.

The remainder of the paper is organized as follows: In Sec.~\ref{sec:layer}, we review the basic signatures of the extended current sheets in the moderate guide field regime. The simulation set-up and profiles from 2D and 3D are presented in Sec.~\ref{sec:ff}. An analysis of the fluctuations in the 3D simulation, including lower hybrid drift and resonant electron firehose modes, are contained in Sec.~\ref{sec:fluct}, and a Summary concludes.

\section{Embedded Electron Current Layer Regime}
\label{sec:layer}

For context, we review the features of the reconnection regime with extended electron current sheets embedded within the exhaust [see Figs.~\ref{fig:comp}(a-b) for examples]. The existence of these current sheets is closely tied to the development of electron pressure anisotropy with differing components of the electron pressure tensor $\pz$ along the local magnetic field and $\pp$ perpendicular to the field. Although this type of anisotropy does not break the electron frozen-in condition \cite{vasyliunas:1975} and enable reconnection in itself, the electron pressure anisotropy regulates the electron current profiles in the reconnection region \cite{hesse:2008,le:2010grl,ohia:2012,le:2013}.

The main mechanism that causes the electron pressure to become anisotropic is electron trapping in a localized parallel electric field structure \cite{egedal:2005,egedal:2008jgr,chen:2008jgr,egedal:2013pop}. This parallel field is an ambipolar electric field that is generated to maintain quasi-neutrality in the reconnection region. It develops within the ion diffusion region, on scales where the ions are essentially unmagnetized, while the electrons continute to follow adiabatic magnetized orbits.

When the bounce frequency of electrons trapped in the parallel electric field's effective potential is faster than the other dynamical time scales, a set of equations of state for $\pz$ and $\pp$ may be derived \cite{le:2009,le:2010pop}. These equations of state asymptotically reduce to the well-known double adiabatic Chew-Goldberger-Low (CGL) scalings \cite{chew:1956, cassak:2015} with $\pz\propto n^3/B^2$ and $\pp\propto nB$ in the limit of a large fraction of trapped electrons. For typical reconnection exhausts, where the magnetic field strength is weaker than upstream and the plasma density is enhanced, the adiabatic compression of the electrons results in a pressure or temperature anisotropy with $\pz>\pp$.

The spatial extent of the electron pressure anisotropy depends on the guide magnetic field. In purely anti-parallel reconnection, the magnetic field goes to zero at the X-line, and the magnetic field is weak in a large region surrounding the X-line. The electron particle orbits are therefore not magnetized, and chaotic electron orbits lead to a nearly isotropic electron pressure tensor in the reconnection outflow. With the addition of a guide magnetic field, however, the electron pressure may remain anisotropic into the reconnection exhaust. The condition for this to occur is roughly that the guide magnetic field is sufficiently strong to keep the ratio $\rho_e/\kappa_B\lesssim0.6$, where $\rho_e$ is the electron Larmor radius and $\kappa_B$ is the radius of curvature of the magnetic field lines \cite{buchner:1989,le:2013,zenitani:2017}.
 
When there is a guide magnetic field strong enough to magnetize the electron orbits, the electron pressure anisotropy may then support a diamagnetic current perpendicular to the local magnetic field. By balancing the $\bf{J\times B}$ force on the electrons with the divergence of the anisotropic electron pressure tensor, we find an additional current driven by the pressure anisotropy of the form ${\bf{J_{e\perp}}}\sim [(\pz-\pp)/B]\bhat\times{\bf{K}}$ \cite{le:2014}. Here, ${\bf{K}}=(\bhat\cdot\nabla)\bhat$ ($\bhat$ is the unit vector along the local magnetic field direction) is the local magnetic field line curvature that signals the presence of a magnetic "tension" force. The divergence of the anisotropic pressure tensor thus balances the magnetic tension force on the current-carrying electrons. This diamagnetic current can form an extended sheet, and it is not confined to electron kinetic scales. Formally, if the pressure anisotropy is great enough to reach the fluid firehose instability threshold $F_e=\mu_0 (\pz - \pp)/B^2=1$, it can support an infinitely long 1D current sheet in the presence of a normal (reconnected) field component \cite{cowley:1978}. This regime with extended current layers embedded in the exhaust (Regime 3 of Ref.~\cite{le:2013}) forms for a range of moderate guide magnetic fields that are at once strong enough to magnetize the electron orbits and yet weak enough that the electron firehose parameter may reach a large fraction $F_e=\mu_0 (\pz - \pp)/B^2\gtrsim0.5$. The long length of the electron current sheets, which can span tens of ion inertial lengths and are limited in simulations only by the size of the simulation domain \cite{ohia:2012,le:2014,le:2016hybrid}, suggests they should be readily observable by spacecraft and in very weakly collisional laboratory experiments \cite{le:2015}. This assumes, however, that the current sheets are stable enough to persist for observable time scales. We examine this question of stability in the following. 

\section{Simulations of a Force-Free Sheet}
\label{sec:ff}

\begin{figure}
\includegraphics[width = 12.0cm]{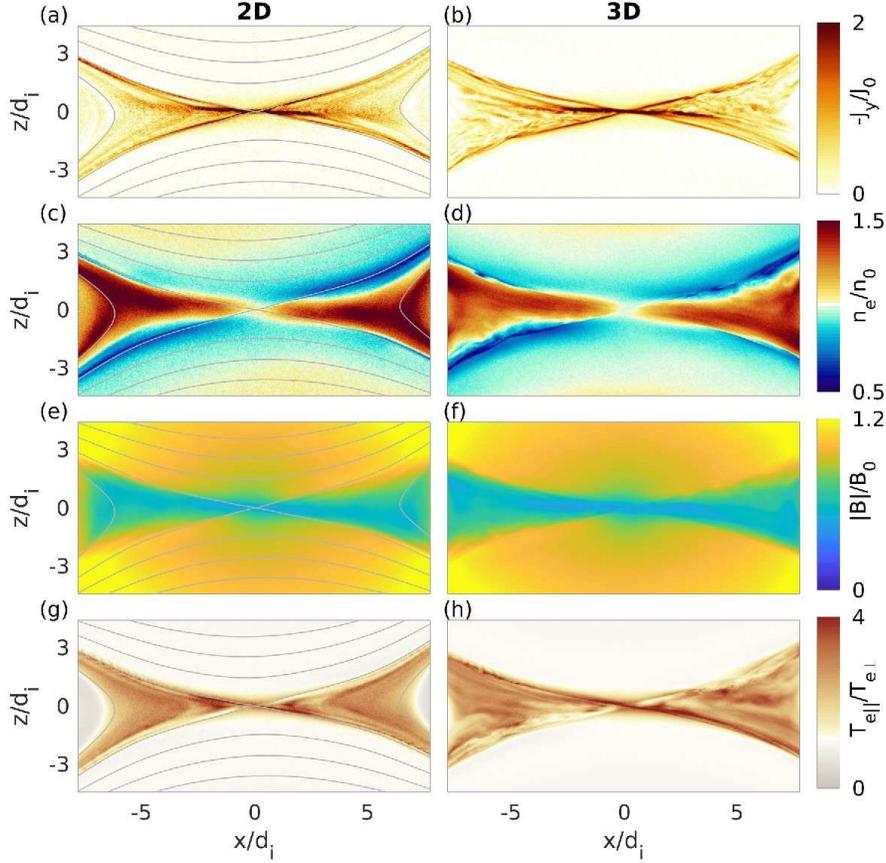}
\caption{Comparison of the reconnecting current sheet from (left) a 2D run and (right) a slice from the 3D simulation. (a-b) The out-of-plane current density contains an extended layer in the exhaust. The extended current sheet is supported by electron pressure anisotropy which develops in the (c-d) increased plasma density and (e-f) decreased magnetic field strength of the exhaust. The (g-h) electron pressure anisotropy peaks around $T_{e\parallel}/T_{e\perp}\sim4$. Gray lines in the plots from the 2D runs on the left are in-plane magnetic field lines. \label{fig:comp}}
\end{figure}

To study embedded current sheets supported by electron anisotropy in reconnection exhausts, we perform fully kinetic particle-in-cell simulations using the code VPIC \cite{bowers:2008} beginning from a force-free, tearing-unstable current sheet. The resulting reconnection exhaust depends mainly on the upstream plasma conditions, and it is very similar to the more commonly considered Harris sheet configuration once reconnection saturates and upstream plasma is advected into the reconnection region. Specifically, the initial magnetic field has the form:
\begin{eqnarray}
B_x(z) &=& B_0\tanh(z/\lambda)\\
B_y(z) &=& \sqrt{B_g^2 + B_0^2- B_x(z)^2}\\
B_z(z) &=& 0
\label{eq:bff}
\end{eqnarray}
and the plasma has a uniform density $n_0$. The electron and ion temperatures are uniform, and they are initialized with $\beta_e = 2\mu_0 n_0 T_e/B_0^2 = 0.1$ and $T_i/T_e=5$. We set $\lambda = 1 d_i = \sqrt{m_ic^2/\epsilon_0 n_0 e^2}$, $\omega_{pe}/\omega_{ce}=1$, and $B_g/B_0=0.3$. The strength of the guide field $B_g/B_0$ and the electron beta $\beta_e$ are particularly important for capturing the regime with embedded electron current sheets \cite{le:2013}.  Reconnection with a single dominant X-line is seeded with a magnetic perturbation of the form
\begin{eqnarray}
\delta B_x(x,z) &=& -\delta B (L_x/2L_z)\cos(2\pi x/L_x)\sin(\pi z/L_z)\\
\delta B_z(x,z) &=&  \delta B \sin(2\pi x/L_x)\cos(\pi z/L_z)
\label{eq:bpert}
\end{eqnarray}
with $\delta B/B_0 = 0.01$.

The numerical particles are sampled from Maxwellian distributions in velocity space, with the electron population drifting to carry the parallel current consistent with the sheared magnetic field profile. We use an ion-to-electron mass ratio $m_i/m_e=300$. Note that a relatively high mass ratio is necessary to produce the correct electron trapping physics and anisotropy in this regime \cite{le:2013}. While at this reduced mass ratio the peak electron temperature anisotropy is not quite as strong as predicted by the adiabatic theory \cite{le:2009}, the pressure anisotropy is nevertheless sufficient to generate an extended electron current sheet embedded within the reconnection exhaust. In order to make the simulation computationally feasible, we employ a reduced frequency ratio of $\omega_{pe}/\omega_{ce}=1$, whereas values of $>10$ are more typical in the magnetosphere. This results in an artificially high electron thermal speed of $v_{the}/c =\sqrt{T_e/m_ec^2}\sim0.3$. While this can neglect Debye-scale fluctuations \cite{jara:2014}, the low value of $\omega_{pe}/\omega_{ce}$ should not greatly alter the meso-scale (roughly ion inertial-scale) physics we focus on here compared to realistic magnetospheric regimes.

The 3D simulation is performed in a numerical domain of size $L_x \times L_y \times L_z = 20d_i \times 20d_i \times 20d_i$ with $1792$ cells in each direction and 150 particles per cell per species ($\sim$ 1.7 trillion numerical particles total). A 2D run was also performed with a single cell in the $y$ direction, but with otherwise identical numerical parameters. The simulations use open boundary conditions \cite{daughton:2006} in the inflow ($z$) and outflow ($x$) directions, and the 3D simulation is periodic in the out-of-plane ($y$) direction. We consider data from time $t\omega_{ci}=51$ when the reconnection exhaust has fully developed.

The reconnection exhausts from the 2D and the 3D simulations are illustrated in Fig.~\ref{fig:comp}, which compares data from the 2D run to a slice in the reconnection plane of the 3D run. The characteristic signature of this regime is the extended current sheet embedded within the reconnection exhaust, which is visible in the out-of-plane current density $J_y$ plotted in Figs.~\ref{fig:comp} (a,b). The current sheet extends to $\sim 5d_i$ from the X-line in both 2D and 3D. As noted above, this distance is limited mainly by the size of the numerical domain, and it is not confined to electron kinetic scales. Again, the embedded electron current layers of this regime are supported by electron pressure anisotropy with $\pz>\pp$. The electron temperature anisotropy [plotted in Figs.~\ref{fig:comp} (g,h)] results from the adiabatic compression of trapped electrons \cite{le:2009,egedal:2013pop} as the plasma density increases [see Figs.~\ref{fig:comp} (c,d)] from the inflow to the reconnection exhaust, while the magnetic field strength decreases [see Figs.~\ref{fig:comp} (e,f)]. 

\begin{figure}
\includegraphics[width = 16.0cm]{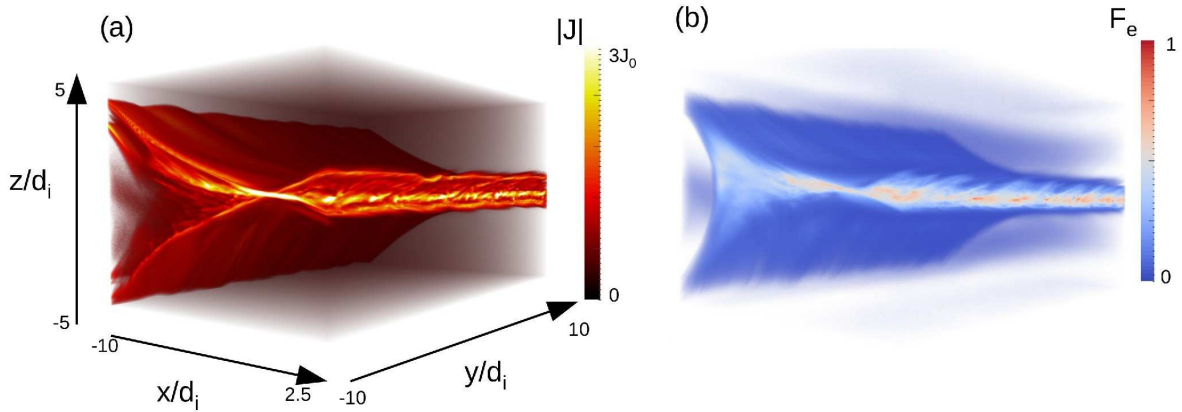}
\caption{Volume renderings from the 3D fully kinetic PIC simulation. Only a subset of the full domain (a cube of side length $20 d_i$) is imaged. (a) Absolute value of the total plasma current density, normalized to $J_0 = n_0ev_{A0}$. The extended current sheet embedded in the exhaust is supported by (b) electron pressure anisotropy with a firehose parameter $F_e = \mu_0(\pz-\pp)/B^2 > 0.5$. \label{fig:3d}}
\end{figure}

The main features of the exhaust in this regime from the 3D simulation are also illustrated in Fig.~\ref{fig:3d}. In panel (a), the absolute value of the total plasma current density is plotted, and it contains a sheet of enhanced current near the X-line corresponding to the extended layer visible in Fig.~\ref{fig:comp}(b). Note that the 3D numerical domain is truncated in the figure to highlight the features of the embedded electron current sheet. In this regime, the electron pressure anisotropy must approach the fluid firehose condition $F_e = \mu_0(\pz-\pp)/B^2\sim1$, and the parameter $F_e$ is rendered in Fig.~\ref{fig:3d}(b). As expected, it reaches $F_e>0.5$ in large sections of the exhaust and is peaked in the region where the embedded electron current sheet resides.

\section{Fluctuations in 3D}
\label{sec:fluct}

\begin{figure}
\includegraphics[width = 18.0cm]{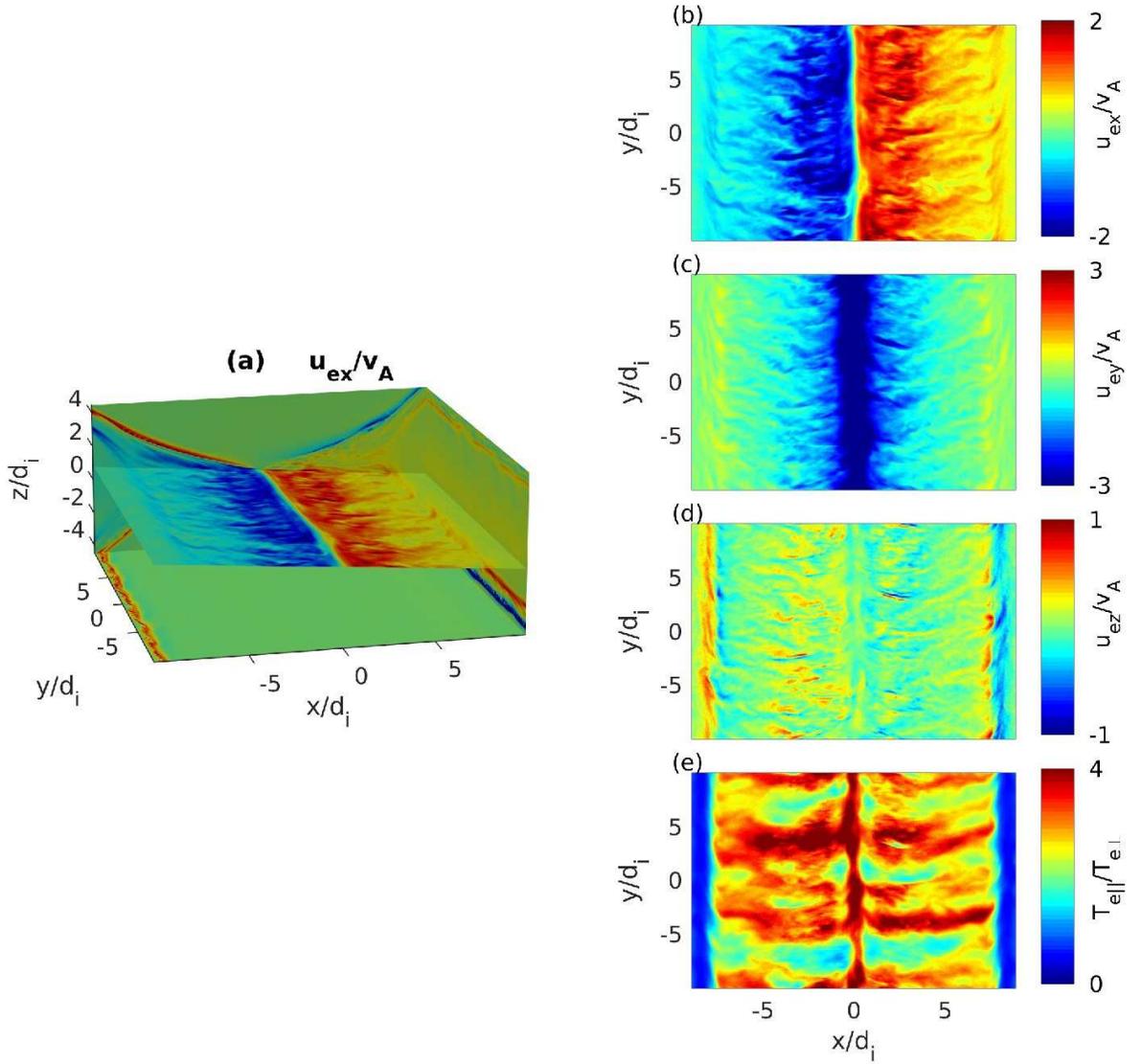}
\caption{Plotted are slices in a plane mainly in the $x-y$ directions, but tilted by $4.5^\circ$ about the $y$ axis to align with the extended current sheet. See panel (a) for a visualization of the location of the cut plane. Plotted are (b-d) the components (in the original, un-tilted simulation frame) of the bulk electron fluid velocity and (e) the electron temperature anisotropy $T_{e\parallel}/T_{e\perp}$. \label{fig:ut}}
\end{figure}

While the general characteristics of the electron current sheet are similar in 2D and 3D, the 3D geometry permits the development of a range of instabilities with finite wave numbers $k_y$ in the out-of-plane direction. In this section, we analyze the variations of the current sheet in the out-of-plane direction and identify the most likely modes that generate fluctuations in 3D that are absent in 2D. To illustrate the variations in the out-of-plane ($y$) direction, we plot slices through the center of the current sheet spanning mainly the $x-y$ plane of the simulation in Fig.~\ref{fig:ut}. The plane is tilted $\sim4.5^\circ$ about the $y$ axis to align with the electron current sheet embedded in the exhaust [see Fig.~\ref{fig:ut}(a)]. The panels in Figs.~\ref{fig:ut}(b-d) show the components of the electron flow velocity, and relatively strong fluctuations over a range of wavelengths are visible in each component.  

\begin{figure}
\includegraphics[width = 15.0cm]{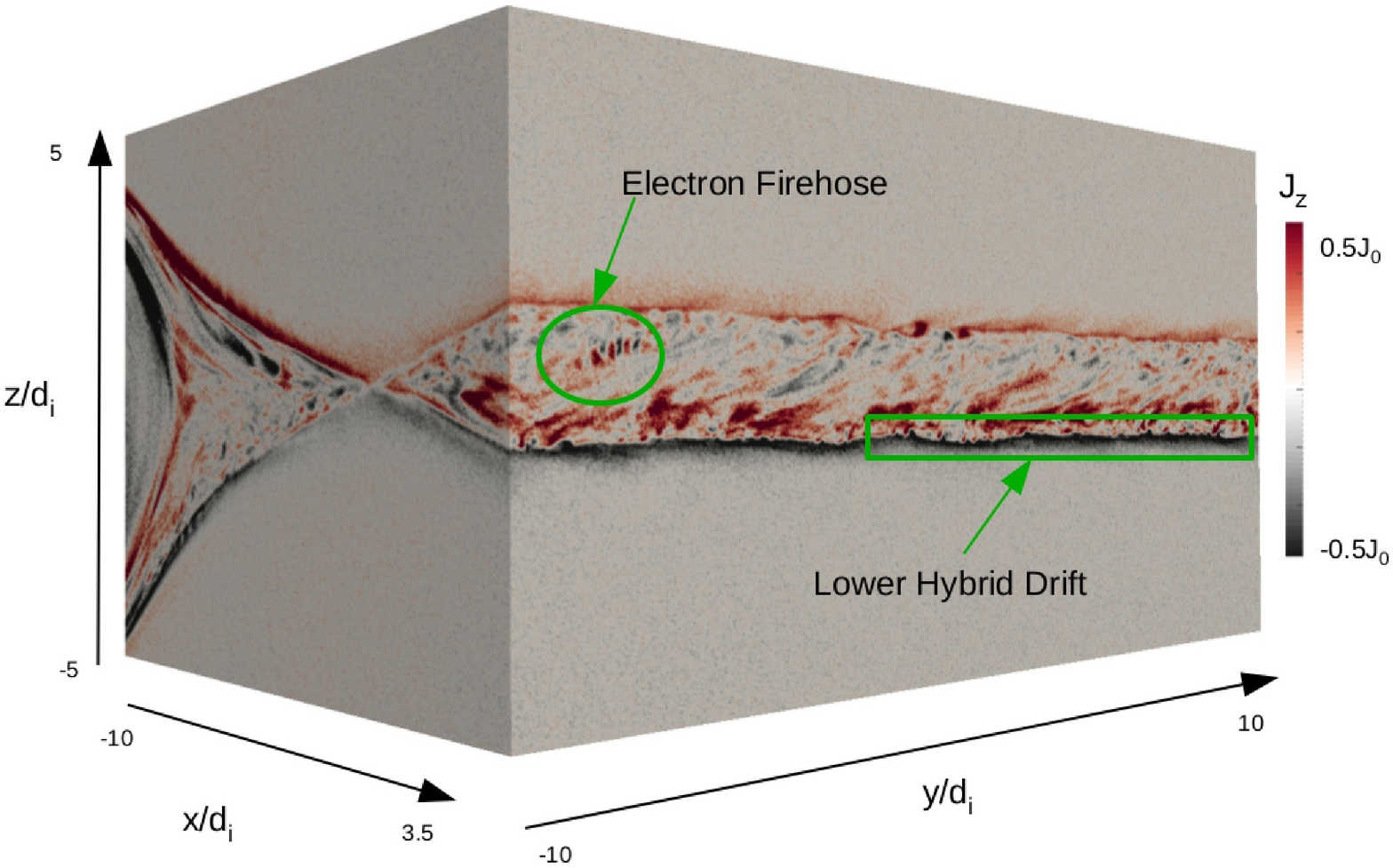}
\caption{Surface plots of the current density $J_z$ from the simulation. The fluctuations in $J_z$ highlight the mode structures of the resonant electron firehose instability within the exhaust and the lower hybrid drift modes that develop mainly along the separatrices. The electron firehose mode develops in a large number of flux tubes scattered throughout the exhaust at different times depending on the local plasma conditions (one example is circled). The lower hybrid drift mode develops along the full extent of the separatrices (an example region is boxed). \label{fig:fhlhdi}}
\end{figure}

One mode with finite $k_y$ present in the 3D simulation is the lower hybrid drift instability (LHDI) \cite{davidson:1975}. LHDI develops even in a standard Harris sheet \cite{daughton:2003}, and it has been observed in a number of 3D reconnection simulations \cite{pritchett:2001b,zeiler:2002,ricci:2004,roytershteyn:2012,price:2016,le:2017,le:2018drift} as well as observations and experiments \cite{carter:2001,bale:2002,zhou:2009,graham:2017}. LHDI is driven by diamagnetic currents at steep density gradients. While the force-free current sheet of our simulation initially has no density gradients, a non-uniform density profile develops self-consistently as reconnection proceeds [see Figs.~\ref{fig:comp}(c-d)]. The steep density gradients along the magnetic separatrices are particularly susceptible to the short wavelength ($k\rho_e\sim0.5)$) electrostatic LHDI, and fluctuations with this wavelength with ${\bf{k\cdot B}}\sim0$ are observed along the separatrix surfaces. To illustrate the location of the LHDI across the separatrices, a 3D view of the current density $J_z$ is plotted in Fig.~\ref{fig:fhlhdi}, and the location of some of the LHDI modes is indicated in a green box. There are also longer wavelength fluctuations within the exhaust that may be the electromagnetic LHDI mode ($k(\rho_e\rho_i)^{1/2}\sim1$) \cite{daughton:2003}, although this mode is largely suppressed by the presence of a guide magnetic field. Interestingly, the longer wavelength fluctuations are accompanied by fluctuations in the electron temperature anisotropy [see Fig.~\ref{fig:ut}(e)], which could result from compressive non-linearities that modulate the local plasma density and magnetic field strength. Because the lower hybrid drift instability is here very similar to previous studies, we do not diagnose its properties in the simulation in detail. 

We devote a more in-depth analysis to another mode in this reconnection regime that depends on the electron temperature anisotropy. There is a relatively short wavelength ($kd_e\sim1$) electromagnetic mode within the exhaust itself, rather than being driven along the separatrices. Its structure along the mid-plane shows most prominently in the electron flow component $u_{ez}$ in Fig.~\ref{fig:ut}(d). As described below, based on its properties, the mode is most likely the resonant oblique electron firehose instability \cite{li:2000, gary:2003}. This mode is suppressed in 2D simulations and depends on the presence of electron temperature anisotropy (and therefore a relatively high mass ratio) in the reconnection exhaust, and it has therefore not previously been studied in reconnection simulations. The resonant electron firehose mode is localized within flux tubes scattered throughout the exhaust of the simulation, as seen in Fig.~\ref{fig:fhlhdi}, where one such patch of short wavelength electromagnetic fluctuations is circled in green. The short wavelength modes develop in different flux tubes over time as the local plasma conditions vary, and there are patches or bursts of the mode visible in multiple flux tubes at any given time once the electron pressure anisotropy has developed.

\subsection{Oblique Resonant Electron Firehose Mode}
\label{subsec:fh}

\begin{figure}
\includegraphics[width = 7.0cm]{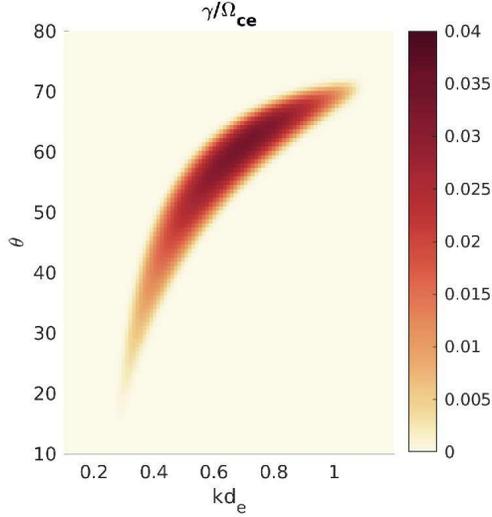}
\caption{Growth rate of the purely growing resonant oblique electron firehose instability for local parameters similar to the exhaust of the reconection simulations: $m_i/m_e=300$, $\beta_{e\parallel} = 2$, $T_{e\parallel}/T_{e\perp}=4$, $\omega_{pe}/\omega_{ce}=1.7$, $T_i/T_{e\parallel}=2$ (no ion temperature anisotropy is included). Normalized to electron scales, the wave vectors and growth rates are relatively insensitive to the ion-to-electron mass ratio. For these plasma parameters, there is an additional growing mode with a real frequency $\omega\sim 2\omega_{ce}$ for parallel propagation ($\theta=0^\circ$) that is not plotted here. \label{fig:fhdisp}}
\end{figure}

Here, we discuss the basic properties of the oblique resonant electron firehose instability \cite{li:2000,gary:2003}, and note that fluctuations observed in the exhaust of our 3D reconnection simulation carry these signatures. The resonant electron firehose instability is a purely growing mode driven solely by the pressure anisotropy of the electrons with $\pz>\pp$. In Fig.~\ref{fig:fhdisp}, we plot the growth rate of the oblique resonant electron firehose instability as a function of wave number $k$ and angle $\theta$ between the wave vector and the magnetic field for parameters typical of the exhaust in the  reconnection simulations. Note that the growth rate is computed for the simplified case of a uniform plasma without ion temperature anisotropy and a bi-Maxwellian electron velocity space distribution. While the growth rates plotted are for the mass ratio used in the simulations ($m_i/m_e=300$), this mode depends primarily on the electrons, and the growth rates and wave numbers normalized to electron scales are nearly identical for the physical mass proton-to-electron mass ratio. The wave numbers with $kd_e\lesssim 1$ are consistent with the 3D simulation. And because of the fast growth rates of $\gamma\sim0.03\omega_{ce}\sim 10\omega_{ci}$, which imply 10 $e$-foldings in an Alfven transit time across 1$d_i$, these modes would quickly saturate on the time scales (10s of ion cyclotron times) of the simulation. We note that for the plasma parameters of Fig.~\ref{fig:fhdisp}, there is an additional unstable electromagnetic mode with $kd_e\sim1$ that propagates parallel to the magnetic field with a real frequency $\omega\sim 2\omega_{ce}$. This mode, however, has a substantially lower growth rate of $\gamma\sim0.3\omega_{ci}$.

One difficulty in definitively identifying this mode in the simulation is that the peak growth occurs at a wave vector that makes an oblique angle of $\sim 60^\circ$ with the magnetic field. In the 3D simulation, however, the electromagnetic fluctuations with $kd_e\lesssim 1$ that we associate with the electron firehose mode are localized to small flux tubes in the exhaust (again see Fig.~\ref{fig:fhlhdi}). The plane wave assumption of the simplest kinetic theories of the mode is therefore not well-satisfied in the simulation, and it is difficult to compute an effective direction of the phase planes in the 3D run. The oblique nature of the resonant electron firehose mode, however, does  explain why it does not develop in the 2D simulation even though the temperature anisotropy stability threshold is exceeded. In 2D simulations where finite $k_y$ is precluded, the angle $\theta$ between a mode wave vector and the magnetic field satisfies $\theta>\arccos(|\bf{B_{xz}}|/|\bf{B}|)$, where $\bf{B}_{xz}$ is the in-plane or poloidal magnetic field and $\bf{B}$ is the total magnetic field including the out-of-plane guide field $B_y$. In the 2D simulation, the guide field is the largest field component over much of the reconnection exhaust, and any mode will have $\theta\gtrsim70^\circ$ over a large region. This geometric effect prevents growth of the oblique electron firehose mode in 2D.

\begin{figure}
\includegraphics[width = 8.0cm]{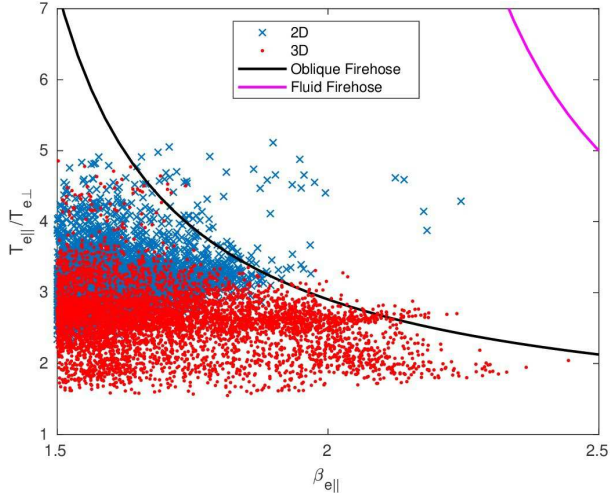}
\caption{Scatter plot of points with $\beta_{e\parallel}>1.5$ in $\beta_{e\parallel}$---$T_{e\parallel}/T_{e\perp}$ space from the reconnection planes plotted in Fig.~\ref{fig:comp}. The magenta curve is the fluid firehose condition $F_e = \mu_0(p_{e\parallel}-p_{e\perp})/B^2 = 1$, and the black curve is the approximate oblique resonant firehose stability condition $T_{e\parallel}/T_{e\perp} = (1-1.29/\beta_{e\parallel}^{0.97})^{-1}$ \cite{gary:2003}. Several points in the exhaust of the 2D run lie well above the resonant firehose stability threshold. \label{fig:fhcomp}}
\end{figure}

As additional evidence of the presence of the resonant oblique electron firehose instability in our 3D simulation, a scatter plot is presented in Fig.~\ref{fig:fhcomp}. Here, the data from the reconnection planes in 2D and 3D from Fig.~\ref{fig:comp} are plotted as points in $\beta_{e\parallel}$---$T_{e\parallel}/T_{e\perp}$ space, including only those points with $\beta_{e\parallel}>1.5$. An approximate form of the stability condition for the oblique electron firehose mode is plotted (black curve), along with the stability threshold for the fluid firehose mode with $F_e = \mu_0(p_{e\parallel}-p_{e\perp})/B^2 = 1$ (magenta curve). While several points of the exhaust in the 2D simulation lie well above the oblique electron firehose threshold, the data from the 3D simulation remain close to the marginal stability curve. This would be expected if the oblique firehose mode, which rapidly isotropizes the electron pressure tensor when it does develop \cite{gary:2003, hellinger:2014}, acts to regulate the electron temperature anisotropy.  

\subsection{Contribution of Fluctuations in the Ohm's Law}
\label{subsec:ohm}

Lastly, we consider whether the fluctuations with finite $k_y$, including the electromagnetic oblique firehose mode, contribute to the Ohm's Law or electron momentum balance equation. Fluctuations such as the lower hybrid drift instability \cite{davidson:1975,huba:1977} have long been considered as possible sources of "anomalous" resistivity that could enhance the reconnection rate in very weakly collisional plasmas. The anomalous terms in the Ohm's Law include anomalous resistivity that results from correlated fluctuations $\delta n$ in the electron density and $\delta E_y$ of the out-of-plane electric field, as well as "anomalous viscosity" \cite{che:2011,price:2016} that is related to the Lorentz force on the electron fluid from correlations in the fluctuating current density and magnetic field. In simulations, the anomalous terms are quantified by separating out the mean $\overline{Q}$ and fluctuating $\delta Q$ components of any quantity $Q = \overline{Q} + \delta Q$, which may be a plasma fluid moment or electromagnetic field component. The average is most often calculated by averaging over the out-of-plane symmetry direction. The $y$-averaging may not be a good measure if there is long wavelength kinking of the current sheet. While time-averaging and time-averaging followed by integration along individual magnetic field lines have also been explored \cite{le:2018drift}, these procedures tend to give similar results in the absence of current sheet kinking. 

\begin{figure}
\includegraphics[width = 14.0cm]{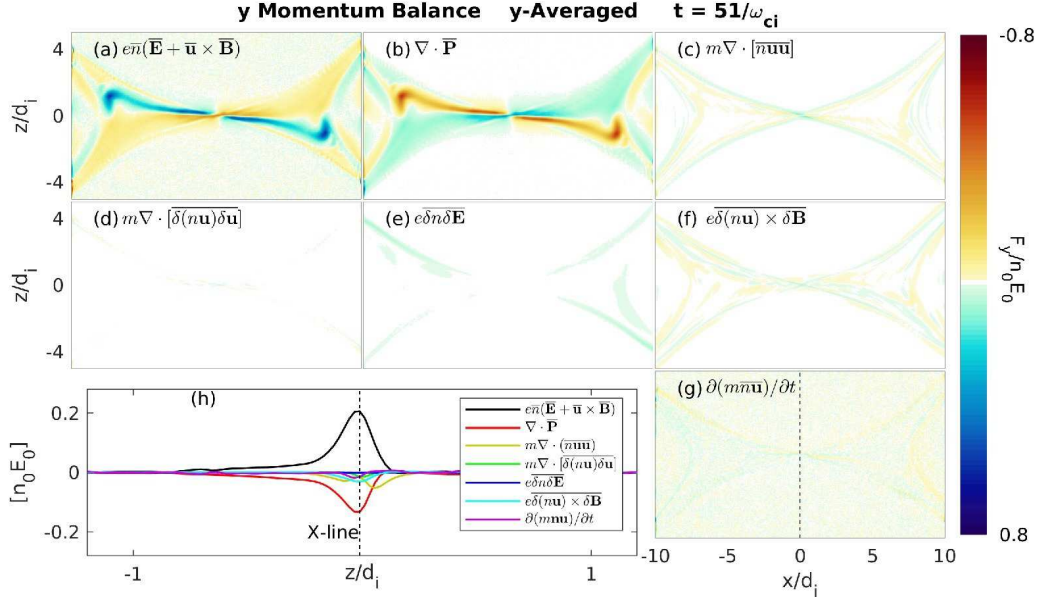}
\caption{Ohm's law with fluctuations averaged on the out-of-plane ($y$) direction. The fluctuations in the exhaust, including the anisotropy-driven electromagnetic modes, do not make a significant contribution to the non-ideal reconnection electric field. The (a) non-ideal field is essentially balanced by (b) the divergence of the electron pressure tensor throughout the domain. The (d-f) anomalous contributions are relatively small. \label{fig:ohm}}
\end{figure}

For this simulation, we applied the $y$-averaging procedure to measure the importance of the fluctuations in the Ohm's Law. The results of this numerical diagnostic are plotted in Fig.~\ref{fig:ohm}. Similar to the results for asymmetric reconnection \cite{roytershteyn:2012,le:2017,le:2018drift}, where the finite $k_y$ drift fluctuations are even more strongly driven, we find that the anomalous terms resulting from correlated fluctuations are small [see Figs.~\ref{fig:ohm}(d-f)]. Rather, the non-ideal electric field in Fig.~\ref{fig:ohm}(a) is almost entirely balanced by the divergence of the electron pressure tensor in Fig.~\ref{fig:ohm}(b). The 3D Ohm's Law through the X-line, as plotted in Fig.~\ref{fig:ohm}(h), is thus very similar to the 2D picture.

\section{Summary}

We studied the 3D stability of extended electron current sheets that form in the exhaust of reconnection sites for a range of guide magnetic fields, typically around $B_g/B_0\sim$0.15---0.5. These current sheets, in "Regime 3" of Ref.~\cite{le:2013}, develop when the pressure anisotropy of the electrons approaches the fluid firehose threshold $F_e = \mu_0(p_{e\parallel}-p_{e\perp})/B^2 \sim 1$. In both 2D and 3D simulations with parameters selected to fall into this regime, an extended layer of electron current developed in the exhaust supported by pressure anisotropy with $\pz>\pp$. In the 3D simulation, a broad spectrum of fluctuations developed that modulated the current density, albeit without destroying the basic structure of the current sheet. These fluctuations included lower hybrid drift modes driven by the density gradients along the separatrices, as well as short wavelength electromagnetic modes within the exhaust itself. The electromagnetic modes were identified as most likely the oblique resonant electron firehose mode, which is driven by the electron pressure anisotropy. This mode is capable of scattering the electrons, and it likely regulated the electron pressure tensor to remain near the instability threshold. In contrast to the adiabatic model that describes the generation of the electron pressure anisotropy \cite{le:2009,egedal:2013pop}, the scattering of electrons by the short wavelength firehose fluctuations is for all practical purposes an irreversible heating process. The reduction in pressure anisotropy caused by scattering off the firehose mode, however, was not strong enough to dissipate the extended current sheets. 

Interestingly, the 3D extended current layer did not break apart into a series of flux ropes, as occurs to some current sheets that develop along the separatrices in other regimes \cite{daughton:2011}. This could be because the presence of strong electron pressure anisotropy \cite{karimabadi:2004,quest:2010} near the firehose threshold and also the presence of a normal (reconnected) magnetic field component within the exhaust both strongly suppress the secondary tearing that generates flux ropes. The relative stability of the extended electron current sheets suggests they should be readily observable embedded in reconnection exhausts in both space observations and laboratory experiments.

\begin{acknowledgments}
Work at LANL was supported by the Basic Plasma Science Program from the U.S. Department of Energy, Office of Fusion Energy Science. A.L. received additional support from NASA grant NNH17AE36I. The 3D fully kinetic simulation was performed on Cori (NERSC). Additional simulations were performed on Pleiades provided by NASA's HEC Program and with LANL Institutional Computing resources.
\end{acknowledgments}

\appendix*
\section{Hybrid (Kinetic Ion/Fluid Electron) Modeling}

In previous work \cite{le:2016hybrid}, we used the hybrid (kinetic ion/fluid electron) code H3D \cite{karimabadi:2014} to study current sheets with anisotropic electron pressure in 2D. Here, we show results from 3D calculations similar to the previous 2D calculations. In the code H3D, ions are treated by a  standard particle-in-cell method. The electron fluid enters the dynamics through an Ohm's law for the electric field of the form:
\begin{equation}
{\bf{E}} = -{\bf{u_i\times B}} + \frac{1}{ne}{\bf{J\times B}} -\frac{1}{ne}\nabla \cdot {\mathbb{P}}_e + \eta {\bf{J}} -\eta_H \nabla^2 {\bf{J}}
\label{eq:e}
\end{equation}
where the current density is defined through Ampere's law without displacement current $\mu_0{\bf{J}} = {\bf{\nabla\times B}}$ and with electron inertia neglected. The magnetic field evolves through the usual Faraday's law $\partial{\bf{B}}/\partial t = -{\bf{\nabla}} \times {\bf{E}}$.

The electron fluid model used two different pressure closures. The simpler closer assumes an isotropic electron pressure $p_e$ and an adiabatic equation of state of the form $p_e\propto n^\gamma$. Here, we use $\gamma=1$, corresponding to the isothermal limit. To capture the electron pressure anisotropy associated with trapping \cite{le:2009,egedal:2008jgr,egedal:2013pop,chen:2008jgr}, the equations of state of Ref.~\cite{le:2009} may be implemented in the code:
\begin{equation}
\tilde{p}_{e\parallel}(\tilde{n},\tilde{B}) = \tilde{n}\frac{2}{2+\alpha} + \frac{\pi\tilde{n}^3}{6\tilde{B}^2}\frac{2\alpha}{2\alpha+1}
\label{eq:ppar}
\end{equation}
\begin{equation}
\tilde{p}_{e\perp}(\tilde{n},\tilde{B}) = \tilde{n}\frac{1}{1+\alpha} + \tilde{n}\tilde{B}\frac{\alpha}{\alpha+1}
\label{eq:pperp}
\end{equation}
where $\alpha = \tilde{n}^3/\tilde{B}^2$. For any quantity $Q$, $\tilde{Q} = Q/Q_\infty$, where $Q_{\infty}$ is the upstream value in a given magnetic flux tube. As demonstrated previously \cite{le:2016hybrid}, this hybrid model captures the basic 2D structure of the extended current sheets supported by electron pressure anisotropy.

\begin{figure}
\includegraphics[width = 10cm]{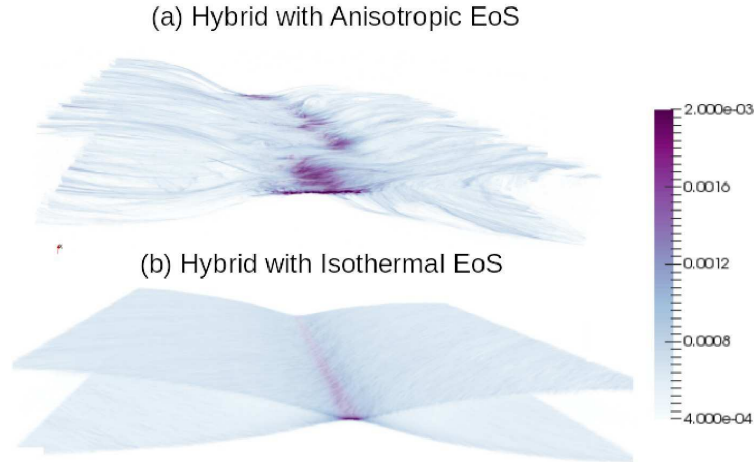}
\caption{Volume rendering of the current density from hybrid simulations of a force-free sheet using similar plasma parameters as the fully kinetic modeling. Simulation in (a) uses anisotropic electron equations of state, while simulation in (b) uses an isothermal electron model. \label{fig:h3d}}
\end{figure}

Volume renderings of the current density from 3D hybrid simulations ($L_x\times L_y\times L_z = 40 \times 40 \times 20 d_i = 512 \times 512 \times 512$ cells) are plotted in Fig.~\ref{fig:h3d} for (a) the anisotropic electron pressure closure \cite{le:2009} and (b) an isothermal electron closure. As in the 2D modeling of Ref.~\cite{le:2016hybrid}, the anisotropic electron pressure with $\pz>\pp$ supports an electron current sheet that extends into the exhaust. By contrast, the plasma current in the model with isothermal electron pressure is confined to a very small region near the X-line. The 3D modeling reveals that while the short current sheet of the isothermal model remains essentially 2D and laminar, the extended current sheet with electron pressure anisotropy develops large, possibly turbulent, fluctuations with variations in the out-of-plane direction. 

While the 3D hybrid model therefore has qualitative features similar to the 3D fully kinetic simulation presented in this paper, there are essential differences. Most importantly, the hybrid model does not support the lower hybrid drift or resonant oblique electron firehose modes that were found in the fully kinetic treatment. The hybrid model will nevertheless include a version of the fluid electron firehose instability when $F_e = \mu_0(p_{e\parallel}-p_{e\perp})/B^2 >1$, although this condition is not reached in the present simulation. Hybrid modeling thus captures the main features of the electron pressure anisotropy-supported current layers, but the nature of the modes that contribute to generating a possibly turbulent reconnection exhaust are not the same as those in a fully kinetic model. Analyzing the microscopic instabilities of the exhaust in a hybrid framework is left for future work.

%\begin{thebibliography}{}
%\bibliographystyle{apsrev}
%\bibliography{coriff}
%\end{thebibliography}

\end{document}